\documentclass[aps,prc,superscriptaddress,showpacs,twocolumn,amsmath]{revtex4}
\usepackage{graphicx}
\usepackage{hyperref}
\usepackage{color}
\begin{document}
\title{Some Characteristic Parameters of Proton from the Bag Model}
\author{Z. G. Tan }\thanks{tanzg@iopp.ccnu.edu.cn}
\affiliation{Department of Electronics and Communication Engineering, Changsha University,Changsha,
410003,P.R.China}
\author{L. Y. Huang}
\affiliation{Department of Electronics and Communication Engineering, Changsha University,Changsha,
410003,P.R.China}
\author{C. B. Yang}
\affiliation{Institute of Particle Physics, Hua-Zhong Normal University, Wuhan, 430079, China}
\begin{abstract}
We treat the mass of a proton as the total static energy which can be separated into two parts that come from
the contribution of quarks and gluons respectively. We adopt the essential of the bag model of hadron to discuss
the structure of a proton and find that the calculated temperature, proton radius, the bag constant are compare
well with QCD results if a proton is a thermal equilibrium system of quarks and gluons.
\end{abstract}

\maketitle

\section{Introduction}\label{sec1}
Exploring proton structure \cite{c1,c2,c3} is still one of  most important subjects for  more profound enhancement of
human knowledge on strong interactions. It is also very helpful for people to search for a new matter state--quark gluon plasma (QGP)
which is the deconfined state of strongly interacting matter. Most theoretical investigations focus on Quantum
Chromodynamics (QCD) \cite{c4}. However,there are a few phenomenological models about nucleon structure and interactions.
The classical string model describes mesons as string segments executing longitudinal expansion and
contraction \cite{c5}. The bag model describes quarks being confined inside a hadron \cite{c6}. In high energy
collisions, the string model represents the process of particle production in the fragmentation of a stretching
string by creating pairs of quark and anti-quark. It worked well \cite{c7} for elementary collisions where
strings can be formed among the few initial partons and break up to form the soft final state hadrons. However,
in relativistic heavy ion collisions (RHIC), there are thousands initial partons. It is intractable to pair
partons and have a string for each pair. Even if the strings are formed, they must be modified by the presence
of many other color charges. The independent fragmentation approach \cite{c7}, though valid for high $Q^2$
partonic processes in the vacuum, cannot explain experimentally observed large $p/\pi$ ratio in central $Au+Au$
collisions at RHIC \cite{c8}. Another possible hadronization mechanism, the deconfined quark (fled out from the
bag) recombination model, has been able to reproduce spectra for almost all stable particles for different
colliding systems. It provides a natural explanation for the baryon/meson ratio and the nuclear suppression
factor observed at RHIC\cite{c9}.

A natural mechanism for quark confinement is given by  the bag model \cite{c6}.  While the bag model has
a few different versions,  we shall in this paper keep the essential characteristics of the phenomenology of
quark confinement. The gluons are mediate bosons which transfer the interactions between quarks . Their effect
can be replaced by the bag pressure which confines the quarks in a hadron. On the other hand, if all interactions
among partons in the bag are neglected, we assume that the partons might be treated as a thermally equilibrated
 system with a given volume. Then properties of a hadron can be investigated and some characteristic
 parameters for the hadron can be obtained in the bag model.

The organization of this paper is as follows. In Sec.\ref{sec2}, we will discuss some features about a thermally
equilibrated QGP system. Then in Sec.\ref{sec3}, we get an estimate of the maximum kinetic energy of a confined
quark in a spherical cavity of radius $R$. Combining the discussions in section II and III by assuming its
origin from the contribution of gluons, the  magnitude of the hadronic bag pressure is discussed in
Sec.\ref{sec4}. The last section is for conclusions and discussions.

\section{Free equilibrated QGP}\label{sec2}
Let us first consider a thermally equilibrated quark-gluon plasma (QGP) system at first. When its temperature $T$ and
volume $V$ are given, the total energy and particle number can be calculated:
\begin{eqnarray}
E&=&\sum_{i=-N_f}^{N_f}\frac{g_i}{(2\pi\hbar)^3}\int f_i(T)p^0\,d\Gamma \nonumber\\
&=&\sum_{i=-N_f}^{N_f}\frac{g_i V}{2\pi^2\hbar^3}\int
f_i(T)p^0|{\bf p}|^2\,dp \,,\label{eev}\\
N&=&\sum_{i=-N_f}^{N_f}\frac{g_i}{(2\pi\hbar)^3}\int f_i(T)\,d\Gamma \nonumber\\
&=&\sum_{i=-N_f}^{N_f}\frac{g_i V}{2\pi^2\hbar^3}\int f_i(T)|{\bf p}|^2\,dp \,,\label{env}
\end{eqnarray}
where $N_f$ is the number of flavors and $g_i=N_cN_s$ is the degeneracy  number for a parton and equals to the
product of quantum numbers of quark's color and spin. $f_i$ is the distribution function which is of Fermi-Dirac for quarks and
Bose-Einstein for gluons
\begin{eqnarray}
&f_i=&\frac{1}{1+e^{(p^0\mp\mu_q)/T}}\qquad \mbox{'-'\, for quark, }\\
&&\hspace{2.8cm} \mbox{'+'\,for anti-quark}\nonumber\\
&f_i=&\frac{1}{e^{p^0/T}-1}\ \qquad \qquad\mbox{ for gluon}\,.
\end{eqnarray}
Here $\mu_q$ is the quark's chemical potential. For the  case when the number density of the quarks
is the same as that of the anti-quarks, $\mu_q=0$.

For a massless quark gas with zero chemical potential $\mu_q=0$, Eqs.(\ref{eev},\ref{env}) can be solved
analytically.  For example for the case with only two flavors, we get the densities for energy and quark
number as ($\hbar=1$)
\begin{eqnarray}
\epsilon=\frac{E}{V}&=&\frac{7}{4}(g_q+g_{\bar{q}})\frac{\pi^2}{30}T^4+ g_g\frac{\pi^2}{30}T^4\nonumber\\
&=&\frac{37}{30}\pi^2T^4 ,\label{fe2}\\
n=\frac{N}{V}&=&\frac{3\zeta(3)}{2\pi^2}(g_q+g_{\bar{q}})T^3+ \frac{g_g\Gamma(3)
\zeta(3)}{2\pi^2}T^3\nonumber\\
&=&\frac{34\times1.202}{\pi^2}T^3 .\label{fn2}
\end{eqnarray}

\begin{figure}
\includegraphics[width=0.45\textwidth]{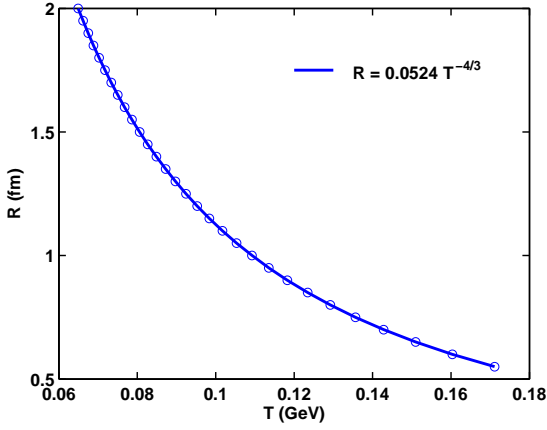}
\caption{The relation between the radius of a proton and its temperature when take its mass as the total energy.
The numerical simulative formula is given on the legend as well.}\label{prt}
\end{figure}

Now if we treat a proton as a free equilibrated QGP system as use the above results,
we can get the relation of the proton radius and temperature. The result is shown in Fig. (\ref{prt}).

From Fig.(\ref{prt}), we can see, the radius of a proton is decreased with its internal temperature. The
numerical simulative formula is
\begin{equation}
RT^{4/3}=0.0524 \label{frt}
\end{equation}
If the temperature is 100 MeV, the corresponding radius of proton is about 1 fm. If a proton is compressed
to have a radius of 0.6 fm, the inner temperature will be about 170 MeV, which is close to the critical temperature,
so that a proton may break and quarks may flee out from the bag.

\section{Quarks confined in a hadron bag}\label{sec3}
Let's give some discussions about quark wave-function from theory. We assume the quarks are confined in a spherical cavity
of radius R, they are free fermions inside it but cannot fly out because all contributions from gluons are
attributed to the bag constant. Then the surface of the sphere bag becomes the
maximum range that quarks can arrive. On the bag boundary where the current of fermion must be zero.
A hadron's wave-function is then product of those for quarks.

The Dirac equation for a free massless fermion in the bag is
\begin{equation}(i\gamma^\mu\partial_\mu-m)\psi=0\quad \mbox{with}\quad m=0, \label{Drc}
\end{equation}
where $\partial_\mu=(p^0,\bf{p})$. We will in this paper use the Dirac representation
\[\gamma^0=\left(\begin{array}{cc}I&0\\0&-I\end{array}\right),\] and
\[\gamma^i=\left(\begin{array}{cc}0&\sigma^i\\-\sigma^i&0\end{array}\right),\]
where $I$ is a $2\times2$ unit matrix and $\sigma^i$ are the Pauli matrices. We write the four-component wave
function for the massless fermion $\psi$ as
\[\psi=\displaystyle{\psi_+\choose\psi_-}\]
where both  $\psi_+$ and $\psi_-$ are two dimensional spinors.  Eq.(\ref{Drc}) becomes
\begin{equation}\left(\begin{array}{cc}p^0&-{\bf \sigma\cdot p}\\{\bf\sigma\cdot p}&-p^0\end{array}\right)
\displaystyle{\psi_+\choose\psi_-}=0\label{Dit}
\end{equation}
The lowest energy solution for the above equation
is the $s_{1/2}$ state given by\cite{c10}
\begin{eqnarray*}\psi_+({\bf
r},t)&=&\mathcal{N}e^{-ip^0t}j_0(p^0r)\chi_+\\
\psi_-({\bf r},t)&=&\mathcal{N}e^{-ip^0t}{\bf \sigma\cdot \hat{r}}j_1(p^0r)\chi_-\,,
\end{eqnarray*}
where $j_l$ is the spherical Bessel function which can be expressed by an elementary function
\begin{equation}j_l(x)=(-1)^lx^l\left(\frac{1}{x}\frac{d}{dx}\right)^l\frac{\sin x}{x},\label{je}
\end{equation}
$\chi_{\pm}$ are two-dimensional spinors, and $\mathcal{N}$ is a normalization constant. The confinement of
the quarks is equivalent to the requirement that the normal component of the vector current
$J_\mu=\bar{\psi}\gamma_\mu\psi$ vanishes at the surface. This condition is the same as the requirement that the
scalar density $\bar{\psi}\psi$ of the quark vanishes at the bag surface $r=R$. This leads to
\[\left.\bar{\psi}\psi\right|_{r=R}=[j_0(p^0R)]^2-{\bf \sigma\cdot\hat{r}\sigma\cdot\hat{r}}[j_1(p^0R)]^2=0\]
or \begin{equation} [j_0(p^0R)]^2-[j_1(p^0R)]^2=0\label{jj}
\end{equation}
From Eq.(\ref{je}), solution of the above equation is given by
\begin{equation}
p^0_mR=2.04,\quad\mbox{or}\quad p^0_m=\frac{2.04}{R} .\label{p0m}
\end{equation}
This result means that in order to keep the bag from being broken, the kinetic energy of any quark can not
larger than $p^0_m$ determined by the radius of the bag.

\section{Bag Pressure}\label{sec4}
Take $p^0_m$ as the upper limit, we can separate the energy of quarks from the total of a proton
\begin{equation}
E_q=\frac{(g_q+g_{\bar{q}})V}{2\pi^2\hbar^3}\int_0^{p^0_m}\frac{p^3dp}{1+e^{p/T(R)}}.
\end{equation}
For simplicity, we have neglected the chemical potential, and treat the quarks as massless.
Then $g_q=g_{\bar{q}}=N_cN_sN_f=3\times2\times2=12$.

The energy from gluon contribution provides the pressure effect directed from the region outside the bag
\begin{equation}
B=\frac{M-E_q}{V}.\label{br}
\end{equation}
Here $M$ is the mass of the hadron.

\begin{figure}
\includegraphics[width=0.45\textwidth]{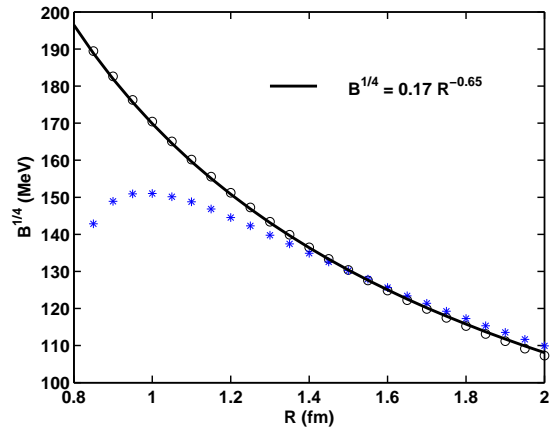}
\caption{The bag pressure change with the radium of proton. The open circle is calculated with Eq.(\ref{br}),
and the line is its numerical simulation results, while the star is with the scenario that three quarks are
surrounded by gluons.}\label{pf2}
\end{figure}

From Eq.(\ref{br}), we can
easily learn that the the bag pressure will change with radius as shown in Fig.(\ref{pf2}).
We also give the numerical formula \begin{equation} B^{1/4}=0.17R^{-0.65}.
\end{equation}
The average kinetic energy of each quark can be calculated as
\begin{equation}
\bar{E}_q=\frac{E_q}{N_q}=\frac{\int_0^{p^0_m}\frac{p^3dp}{1+e^{p/T(R)}}}
{\int_0^{p^0_m}\frac{p^2dp}{1+e^{p/T(R)}}}.
\end{equation}

The energy carried by the valence quarks in a proton is then $3\times\bar{E}_q$.
So that contributions from gluons and sea quarks to the energy is $M-3\bar{E}_q$.
The bag pressure decreases with the radium.
This may be used to explain why the resonance particle (which has large radius thus small bag pressure) is
usually unstable because their bags are more fragile.

\section{Conclusion}
In this work we have discussed some feature of a proton from the bag model. Especially, we got the relations
between temperature and the radius of the proton when a proton is treated as a non-interacting thermal equilibrium
QGP system.  The
bag pressure comes from the contribution of gluons.

\acknowledgments{We are grateful to the financial support from China ChangSha University under grant
No.SF080101. We also thank Prof. A. Bonasera for stimulating discussions and comments.

\end{document}